\documentclass[lettersize,journal]{IEEEtran}
\usepackage{amsmath,amsfonts}
\usepackage{algorithmic}
\usepackage{array}
\usepackage[caption=false,font=small,labelfont=rm,textfont=rm]{subfig}
\usepackage{textcomp}
\usepackage{stfloats}
\usepackage{float}
\usepackage{url}
\usepackage{verbatim}
\usepackage{graphicx}
\usepackage{xcolor}
\usepackage[justification=centering]{caption}
\usepackage{hyperref}
\usepackage{amssymb}
\usepackage{multirow}
\usepackage{booktabs}
\usepackage{tabularx,array}

\def\BibTeX{{\rm B\kern-.05em{\sc i\kern-.025em b}\kern-.08em
    T\kern-.1667em\lower.7ex\hbox{E}\kern-.125emX}}
\usepackage{balance}

\begin{document}

\title{MoPHES:Leveraging on-device LLMs as Agent for Mobile Psychological Health Evaluation and Support}
\author{Xun Wei $^{\dagger*}$\thanks{Xun Wei is with the School of Software Engineering, Jiangxi University of Science and Technology, Nanchang 330044, China (email:xun.wei@jxust.edu.cn).},
\and Pukai Zhou $^\dagger$ \thanks{Pukai Zhou is with the School of Computer Science, Shenzhen University, Shenzhen 518060, China (e-mail: 2510103072@mails.szu.edu.cn).},
\and Zeyu Wang \thanks{Zeyu Wang is with the School of Mathematics and Computer Sciences, Nanchang Univertity, Nanchang 330031, China (email: wangzeyuwangzeyu@email.ncu.edu.cn).}
}

\markboth{Journal of IEEE,~Vol.~18, No.~9, October~2025}%
{MoPHES:Leveraging on-device LLMs as Agent for Mobile Psychological Health Evaluation and Support}

\maketitle

\renewcommand\thefootnote{\fnsymbol{footnote}}
\footnotetext{$\dagger$ Equal Contribution.}
\footnotetext{* Corresponding Author.}

\renewcommand\thefootnote{\arabic{footnote}} 

\begin{abstract}
The 2022 World Mental Health Report calls for global mental health care reform, amid rising prevalence of issues like anxiety and depression that affect nearly one billion people worldwide. Traditional in-person therapy fails to meet this demand, and the situation is worsened by stigma. While general-purpose large language models (LLMs) offer efficiency for AI-driven mental health solutions, they underperform because they lack specialized fine-tuning. Existing LLM-based mental health chatbots can engage in empathetic conversations, but they overlook real-time user mental state assessment which is critical for professional counseling.
This paper proposes {\normalfont\texttt{MoPHES}}, a framework that integrates mental state evaluation, conversational support, and professional treatment recommendations. The agent developed under this framework uses two fine-tuned MiniCPM4-0.5B LLMs: one is fine-tuned on mental health conditions datasets to assess users' mental states and predict the severity of anxiety and depression; the other is fine-tuned on multi-turn dialogues to handle conversations with users. By leveraging insights into users' mental states, our agent provides more tailored support and professional treatment recommendations. Both models are also deployed directly on mobile devices to enhance user convenience and protect user privacy.
Additionally, to evaluate the performance of MoPHES with other LLMs, we develop a benchmark for the automatic evaluation of mental state prediction and multi-turn counseling dialogues, which includes comprehensive evaluation metrics, datasets, and methods.
\footnote{\href{https://github.com/weixun2018/MoPHES}{https://github.com/weixun2018/MoPHES}}
\end{abstract}

\begin{IEEEkeywords}
Mental Health, large language model, intelligent agent, psychological understanding, dialogue system.
\end{IEEEkeywords}

\section{Introduction}
\IEEEPARstart{M}{ental} health issues are emerging as an increasingly severe threat to global public health, with their prevalence rising annually. Approximately one billion people worldwide suffer from psychological disorders, accounting for 13\% of the global population and imposing a heavy disease burden \cite{WHO}. Among these mental disorders, anxiety and depression alone make up nearly 60\%. However, this threat remain significantly underestimated. Over 70\% individuals with such disorders never access effective mental health services, primarily due to low social awareness and stigma surrounding mental illness, particularly in developing countries \cite{WHO}. Furthermore, traditional in-person therapy includes both offline sessions and online consultations with a psychological counselor, but it still struggles to meet the enormous demand. The inadequacy of this approach comes from two key limitations: a shortage of qualified professionals and high service costs. As a result, the challenge of providing high-quality, affordable mental health care remains formidable.

Recently, the Natural Language Processing(NLP) technology has been actively applied to develop AI-powered systems that deliver psychological counseling and treatment guidance. Researchers have primarily focused on three core areas:providing mental disease counseling, enhancing emotional support capability and offering online psychological consultation services. Notably, the advent of LLMs has significantly advanced AI intelligence and fuels enthusiasm of researchers into digital mental health interventions. Subsequently, various studies have been proposed to improve mental health by leveraging LLMs as conversational chatbots or dedicated intelligent agents \cite{digital-com,Counseling,ChatCounselor,Emo-chatbot}. 

However, the direct application of general-purpose LLMs, such as ChatGPT, Claude and Llama, tends to yield underwhelming performance in mental health field. Naturally, an effective method to overcome this limitation is to fine-tune general-purpose LLMs with specialized psychological corpora. Currently, a series of mental health LLMs have been developed, including MeChat \cite{Smile}, PsyLLM \cite{Psy-llm} , SoulChat \cite{SoulChat}, CPsyCoun \cite{CPsyCoun}, EmoLLLM \cite{EmoLLM}, PsycoLLM \cite{PsycoLLM}. Owing to ethics policy and privacy protection, real-world datasets of multi-turn mental counseling dialogues remain extremely scarce. Thus, the key of these studies is to construct a high-quality multi-turn dialogues dataset of psychological counseling, which are typically derived from website-crawled Q\&As or clinical counseling reports. For instance, SMILECHAT \cite{Smile} is synthetic multi-turn dialogues dataset containing 55K samples, generated from single-turn QAs. When carefully fine-tuned on such multi-turn conversions, mental health LLMs exhibit superior performance compared to general-purpose LLMs in terms of content naturalness, emotional empathy and helpfulness.

Despite these advancements, the limitation of the aforementioned chatbots lies in their failure to conduct concurrent assessments of users' mental states during interactions—a core capability for professional psychological counselors \cite{mental_state}. For professional psychological counseling, a counselor typically adopts corresponding psychotherapy methods based on their synchronous evaluation of a user's mental state. For example, symptoms of mild anxiety condition differ from those of severe anxiety condition, and the corresponding treatment approaches also vary: the former may only require appropriate conversational guidance and meditation, while the latter is likely to need medication-based intervention. Consequently, it is necessary to enable LLMs to evaluate users' mental states and deliver symptom-specific interventions. Additionally, in the mobile internet era, the provision of mobile psychological counseling services has become extremely urgent.

In this paper, we propose a novel framework \textbf{MoPHES}, which leverages LLMs as intelligent agent to provide mobile psychological health evaluation and support. The implementation of MoPHES follows three key stages, detailed as follows:
First, we collect single-turn counseling QAs from publicly available sources. For dataset construction, we label users' counseling questions with mental conditions labels; simultaneously, we generate multi-turn dialogues from the single-turn QAs via prompting GPT-4o mini model.
Then we fine-tune two MiniCPM4-0.5B LLMs separately on the mental health conditions dataset and the multi-turn dialogue dataset. The first model, fine-tuned on the mental conditions dataset, focuses on evaluating users' mental states; the second model, fine-tuned on the multi-turn dialogue dataset, engages users in empathetic conversations.
Finally, we deploy the two fine-tuned models on Android-based mobile devices using the llama.cpp framework \cite{llama.cpp}. During interactions, the agent assesses user's mental state and stores the assessment results locally after every 5 dialogue turns. When initiating the next dialogue round, the agent first loads the user's historical mental state records, integrates this information with the user's current input, and then generates a more tailored and context-aware response.

The main contribution of this paper can be summarized as follows:
\begin{itemize}
  \item We propose a new framework for mental health that organically integrates the mental state assessment, conversational support and professional treatment recommendations. By fine-tuning on mental health conditions dataset and multi-turn counseling dialogues, our agent is endowed both predictive and conversational capabilities and more closely mimics the role of a real psychologist.
  \item We demonstrate that “small” LLM with 0.5 billion parameters can achieve remarkable performance in the mental health domain through fine-tuning on specialized corpora. This model can be easily deployed and run on most users' mobile devices, thereby enabling the provision of convenient mental health services while protecting users' privacy.
  \item We develop a psychological benchmark for automatic evaluation of mental state prediction and multi-turn  counseling dialogues, which includes comprehensive evaluation metrics, datasets and methods.
\end{itemize}

\section{Related Work}

Recently, the mental health field has garnered significant attention across numerous studies, largely driven by the impact of the COVID-19 pandemic \cite{Factors,Behavioral,Ensemble,calibration,Graph}. Since the emergence of LLMs, researchers have begun to leverag these models to support mental health efforts, primarily focusing on areas such as mental conditions identification and psychological chatbot development.

MentaLLaMA \cite{MentalLlama} introduced a multi-task and multi-source interpretable mental health instruction
(IMHI) dataset with 105K data samples collected from 10 existing sources covering 8 mental health analysis tasks. Based on IMHI dataset and LLaMA2 \cite{Llama2} foundation models, MentaLLaMA was trained for interpretable mental health analysis on social media.
MentalLLM \cite{MentalLLM} presented a comprehensive evaluation of prompt engineering, few-shot, and finetuning techniques on multiple LLMs in mental health domain. Meanwhile it fine-tuned Alpaca-7B and FLAN-T5-XXL models with seven online social media datasets and demonstrated significant improvement of LLM's capability on multiple mental-health-specific tasks across different datasets simultaneously.

The use of chatbots to support mental health has a well-established history \cite{review}. In the early stages, rule-based systems were the primary approach for building chatbots. These systems employed various therapeutic techniques to guide users through self-help exercises \cite{SERMO}, but their rigid rule-based design inherently limited conversational naturalness \cite{typing-cure} and left them unable to fully understand users' concerns.

The advent of LLMs has since sparked a new wave of interest in the potential of conversational agents for mental health support, such as OpenAI's ChatGPT \cite{chatgpt}. These LLM-powered chatbots, equipped with user-friendly conversational interfaces, have ignited excitement among clinicians about the potential of innovative AI-driven mental health interventions. Designed to enable direct interaction, these chatbots connect with individuals seeking mental health support across diverse platforms, including personal digital companions \cite{digital-com}, online on-demand counseling \cite{Counseling,ChatCounselor} and emotional support services \cite{Emo-chatbot}.
However, platform-based chatbots like ChatGPT often lack empathy. Theny tend to provide repetitive and standardized responses and prioritize offering suggestions over asking follow-up questions or engaging in active listening—shortcomings that prevent them from truly meeting users' emotional needs.

In order to address these limitations and make LLMs more human-like, many studies have been proposed recently to enhance the empathic ability of general-purpose LLMs by fine-tuning them on specialized psychological corpora.
PsyQA \cite{PsyQA} is a high-quality chinese dataset of psychological health support in the form of one question mapping to multiple answers, which is crawled from a Chinese mental health service platform. 
SIMLE \cite{Smile} introduced a technique that prompts ChatGPT to rewrite public single-turn QAs based on PsyQA into multi-turn dialogues to develop specialized dialogue systems for mental health support. It also builds a mental health chatbot MECHAT by fine-tuning ChatGLM2-6B \cite{ChatGLM} with the collected corpora. 
PsyChat \cite{PsyChat} proposed a client-centric dialogue system for mental health support that can predict the client's behaviors, select appropriate counselor strategies, and generate accurate and suitable responses.
Psy-LLM \cite{Psy-llm} also utilized PsyQA dataset to fine-tune PanGu-350M \cite{pangu} model to alleviate the demand for mental health professionals. 
SoulChat \cite{SoulChat} was designed to be more “human-centered” for psychological support by constructing a multi-turn empathetic conversation dataset with more than 2 million samples, in which the input is the multi-turn conversation context, and the target is empathetic responses. 
PsyDT \cite{SoulChat-v2} is the version 2.0 of SoulChat that aims to satisfy the individual needs of clients. It is a novel framework using LLMs to construct the digital twin of psychological counselor with personalized counseling style. The core of PsyDT is the multi-turn dialogues synthesis method, which consists of three components: dynamic one-shot learning from counseling cases, client personality simulation and multi-turn mental health dialogues synthesis. 
CPsyCoun \cite{CPsyCoun} presented a report-based multi-turn dialogue reconstruction and evaluation framework for chinese psychological counseling. It introduced a two-phase method named MEMO2DEMO to construct high-quality dialogues and develop a comprehensive evaluation benchmark for the effective automatic evaluation of multi-turn psychological consultations. 
PsycoLLM \cite{PsycoLLM} was trained on a proposed high-quality psychological dataset, including single-turn QA, multi-turn dialogues and knowledge-based QAs. It also developed a comprehensive psychological benchmark based on authoritative psychological counseling examinations in china, which includes assessments of professional ethics, theoretical proficiency, and case analysis.

Building on the aforementioned foundation, our agent can simultaneously assess users' mental states during conversations. This capability enables it to function more like a psychologist and deliver more professional support. Furthermore, our research provides mobile mental health services by leveraging on-device LLMs, which can be easily deployed on mobile devices and operate seamlessly.

\section{MoPHES}
\begin{figure*}[htbp]
\centering
\includegraphics[width=\textwidth]{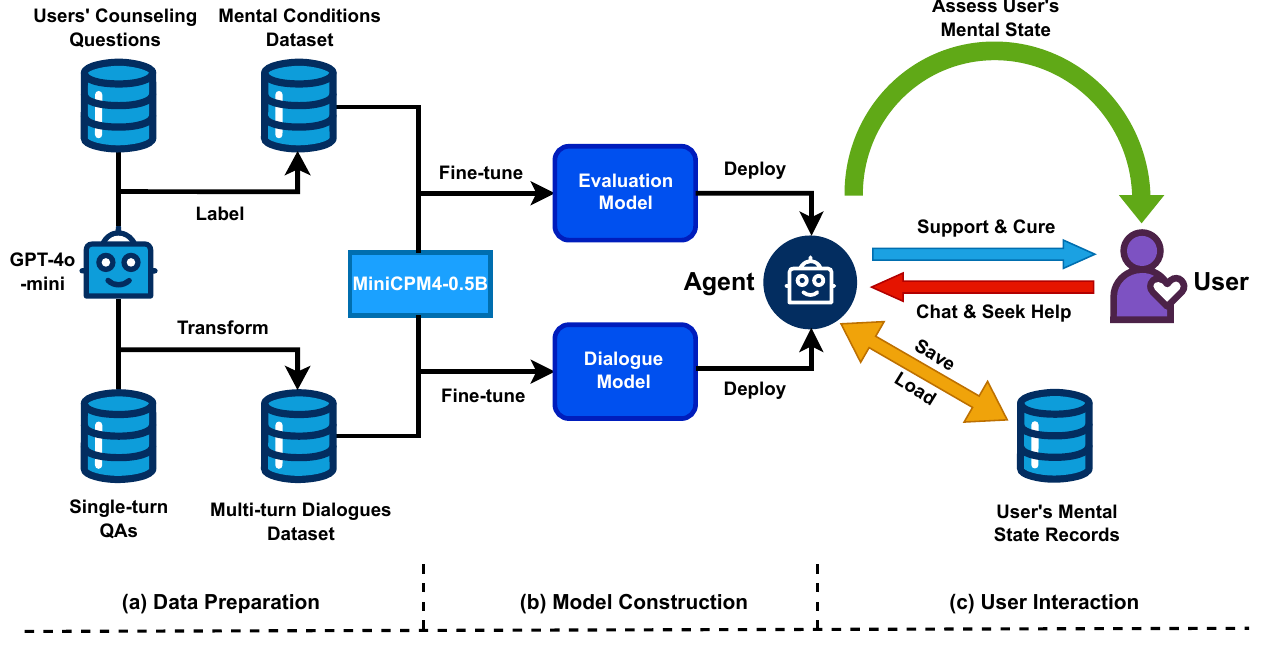}
\caption{Overview of MoPHES. (a) We use GPT-4o-mini to label the mental conditions dataset and transform singile-turn QAs to multi-turn dialogues; (b) We fine-tune the base model MiniCPM4-0.5B on these two datasets and obtain evaluation model and dialogue model respectively, and deploy the models on mobile device to build the agent; (c) The interaction between user and agent: user can chat with agent and seek help via multi-turn conversations, and agent will support and cure user, meanwhile regular assess user's mental state and save them locally.}
\label{fig:overview}
\end{figure*}

In this section, we elaborate on the complete workflow of \textbf{MoPHES}, including datasets preparation, model construction and user interaction. First, we illustrate the data source of public singile-turn QAs and mental conditions dataset, data preprocessing, and demonstrate the techniques that label the mental conditions dataset and transform singile-turn QAs to multi-turn dialogues. Then we introduce the base model adopted in framework. Subsequently, we demonstrate the process of supervised fine-tuning and Low-Rank adaptation. Finally, we show the deployment and usage of our agent. The overview of \textbf{MoPHES} is shown in Fig.~\ref{fig:overview}.

\subsection{Datasets}
Overall, fine-tuning datasets are consists of two parts: subthreshold mental conditions dataset and multi-turn dialogues dataset. Considering the prevalence of mental conditions, we construct this classification dataset with only two conditions: anxiety and depression, and each condition with four levels of severity. Owing to the lack of available chinese dataset on mental conditions, we utilize AI model to label users' counseling questions with severity of anxiety and depression condition. On the other hand, due to the ethics policy and privacy protection, real multi-turn dialogues of mental counseling are exceedingly rare, we have drawn on the methods of our predecessors \cite{Smile} and transform public single-turn QAs to multi-turn dialogues. 

\noindent\textbf{Data Source}

The datasets used in this study were constructed from two publicly available psychological counseling resources: PsyQA \cite{PsyQA} and EmoLLM \cite{EmoLLM}. After preprocessing, the PsyQA provided 81,219 valid single-turn QAs while the EmoLLM dataset contributed 32,333 samples.

By merging two resources, we obtained a total of 113,552 initial QAs. Each sample consists of a user query and a corresponding assistant reply. It is worthnoting that the dataset is entirely in Chinese, all data were collected from open sources, and reformatted for research purposes.

\noindent\textbf{Data Preprocessing}

To ensure the quality and consistency of the dataset, we applied several preprocessing steps. First, we set a length threshold: samples were removed if the user query contained fewer than 50 characters or the assistant reply contained fewer than 100 characters. Next, we used an AI-based filter to detect and remove low-quality or irrelevant responses, which eliminated 41,083 samples. 

Finally, we applied MinHash with Locality-Sensitive Hashing (LSH) to detect near-duplicate pairs at a 70\% similarity threshold, removing around 33,232 duplicates. These steps produced a cleaner and more reliable dataset, which provides a solid foundation for later experiments.

\noindent\textbf{Data Statistic}

\begin{figure}[htbp]
    \centering
    \includegraphics[width=0.45\textwidth]{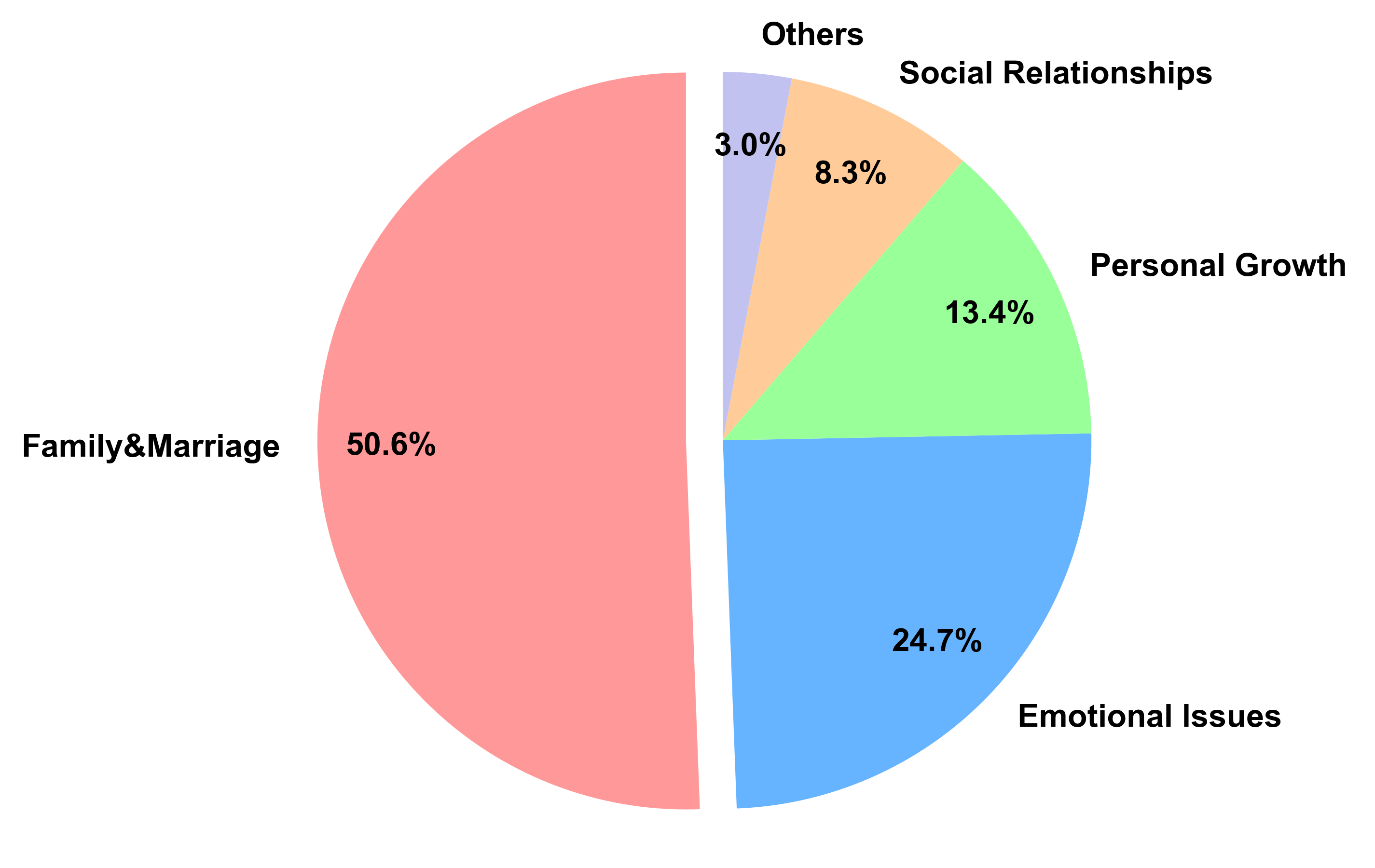}
    \caption{Distribution of counseling topics.}
    \label{fig:training-set}
\end{figure}

After preprocessing and deduplication, we obtained 34,827 single-turn QA pairs. Fig.~\ref{fig:training-set} shows the distribution of counseling topics. The largest proportion is Family and Marriage (50.6\%), followed by Emotional Issues (24.7\%), Personal Growth (13.4\%), Social Relationships (8.3\%), and Others (3.0\%). 

\noindent\textbf{Mental Conditions Labeling}

Firstly, we filtered the previous QAs according to text length: keep the samples which question has more than 200 characters. We retained counseling questions and removed the duplicates using LSH method. Then We prompted GPT-4o-mini to label the users' counseling questions with the severity of depression and anxiety condition respectively (Appendix \ref{sec:prompts}). The severity has four level: minimal, mild, moderate and severe, labeled from 0 to 3 correspondingly. Finally, we obtained 6,046 labeled samples to build mental conditions dataset. Notably, these at-risk mental conditions were labeled based on users' text input instead of clinical diagnosis.

Table~\ref{tab:distribution} shows the distribution of depression and anxiety severity in our dataset. Obviously, most clients suffer from anxiety and depression across a spectrum of severity. Specifically, only 1.3\% clients maintain fully sound mental health while nearly 30 \% clients experienced moderate anxiety and moderate depression. Moreover, the distribution of depression severity is relatively uniform, while anxiety conditions are mainly concentrated in the moderate level.

\begin{table}[h]
\centering
\caption{Distribution of depression and anxiety severity in the mental conditions dataset.}
\label{tab:distribution}
\renewcommand{\arraystretch}{1.25}
\resizebox{\linewidth}{!}{
\begin{tabular}{c|cccc|c}
\toprule
\textbf{Depression\textbackslash Anxiety} & Minimal & Mild & Moderate & Severe & \textbf{Sum} \\
\midrule
Minimal   & 79 & 140 & 44   & 2   & 265 \\
Mild      & 2  & 314 & 1233 & 168 & 1717 \\
Moderate  & 0  & 259 & 1602 & 483 & 2344 \\
Severe    & 4  & 27  & 1186 & 503 & 1720 \\
\hline
\textbf{Sum} & 85 & 740 & 4065 & 1156 & 6046 \\
\bottomrule
\end{tabular}
}
\end{table}

\noindent\textbf{Multi-turn Dialogues Generation}

Intuitively, single-turn QA pairs were expanded into 5-turn dialogues. Typically, an assistant is designed to understand users' concerns and analyse their emotional states in the previous turns of the dialogue, before finally providing targeted treatment suggestions. The meticulously designed prompt (Appendix \ref{sec:prompts}) guided the GPT-4o-mini model to generate concise, natural, and counseling-oriented multi-turn conversations in Chinese. We set a low temperature (0.2) and a maximum length of 350 tokens to ensure stability. Then we use an AI-based filter to remove low-quality cases. This process systematically transformed the rigid single-turn QA pairs into empathy multi-turn dialogues. The statistics of multi-turn dialogues are summarized in Table~\ref{tab:dialogues}.

\begin{table}
  \caption{Statistics of Multi-turn Dialogues Dataset}
  \label{tab:dialogues}
  \renewcommand{\arraystretch}{1.25}
  \begin{center}
    \begin{tabular}{l|c}
    \toprule
    \textbf{Category}              & \textbf{Size}     \\
    \midrule      
    \# Dialogues                    & 34381  \\
    \# Average turns per dialogue   & 5.00   \\
    \# Average tokens per turn      & 76.27  \\
    \# Average tokens per question  & 28.02  \\
    \# Average tokens per answer    & 48.25  \\
    \bottomrule
    \end{tabular}
    \end{center}
\end{table}

We present a case of data preparation that includes mental conditions labeling and multi-turn dialogue generation in Appendix \ref{sec:case}.

\subsection{Backbone Model}
The base model selected is MiniCPM4, a highly efficient large language model designed explicitly for end-side devices \cite{MiniCPM4}. This model is optimized via innovations in four areas: architecture (InfLLM v2, a trainable sparse attention mechanism for long-context processing), training data (UltraClean filtering/generation and UltraChat v2 dataset, enabling good performance with 8T tokens), algorithms (ModelTunnel v2 for pre-training strategy search, and improved post-training methods like chunk-wise rollout and BitCPM), and inference systems (CPM.cu integrating sparse attention, quantization, and speculative sampling). Available in 0.5B and 8B parameter versions, it outperforms similar open-source models across benchmarks, with the 8B variant showing faster long-sequence processing than Qwen3-8B. It also supports diverse applications like survey generation and tool use with model context protocol, demonstrating its broad usability. For higher speed on mobile phone, we adopt the version with 0.5B parameters. 

\subsection{Supervised Fine-tuning and Low-Rank Adaptation}
After the datasets and backbone model were prepared, we adopt a typical technique, supervised fine-tuning (SFT), to enhance the performance of LLMs in the mental health domain. Due to the capability limitations of the selected on-device LLMs, we train two LLMs separately on the mental conditions dataset and the multi-turn dialogue dataset. The LLM fine-tuned on the mental conditions dataset can assess users' mental states and predict the severity of anxiety and depression. Meanwhile, the other LLM, which is fine-tuned on the multi-turn dialogues dataset, is responsible for chatting with users empatheticly.

With the preparation of a labeled dataset \(\mathcal{D} = \{(X^{(i)}, Y^{(i)})\}_{i=1}^N\), where each \(X^{(i)} = (x_1^{(i)}, x_2^{(i)}, ..., x_T^{(i)})\) represents an input sequence and \(Y^{(i)} = (y_1^{(i)}, y_2^{(i)}, ..., y_T^{(i)})\) denotes the corresponding ground-truth generation output of length \(T\). Prior to training, the input-output pairs are tokenized using the model's native tokenizer. During the SFT training process, the model is tasked with maximizing the conditional probability of generating the ground-truth output \(Y^{(i)}\) given the input \(X^{(i)}\), which is formalized by minimizing a cross-entropy loss function. Specifically, the loss \(\mathcal{L}_{\text{SFT}}(\Theta)\) for the model with parameters \(\Theta\) is defined as Equation \eqref{equ:loss}:

\begin{equation}\label{equ:loss}
\mathcal{L}_{\text{SFT}}(\Theta) = -\sum_{i=1}^N \sum_{t=1}^L \log p_\Theta(y_t^{(i)} \mid x_1^{(i)}, ..., x_t^{(i)}), 
\end{equation}
where \(p_\Theta(y_t^{(i)} \mid x_1^{(i)}, ..., x_t^{(i)})\) is the probability of the model predicting the \(t\)-th token of \(Y^{(i)}\) given the preceding token sequence. Optimization is conducted via gradient-based algorithms, where the model parameters \(\Theta\) are iteratively updated using:

\begin{equation}\label{equ:theta}
\Theta_{k+1} = \Theta_k - \eta \cdot \nabla_\Theta \mathcal{L}_{\text{SFT}}(\Theta_k), 
\end{equation}
with \(\eta\) representing the learning rate and \(\nabla_\Theta \mathcal{L}_{\text{SFT}}(\Theta_k)\) the gradient of the loss with respect to \(\Theta\) at the \(k\)-th iteration; this process continues until the validation loss on a held-out dataset stabilizes, ensuring the model adapts to the task without overfitting or catastrophically forgetting pre-trained knowledge.

Low-Rank Adaptation (LoRA) \cite{lora} is a parameter-efficient fine-tuning technique for generative LLMs, addressing full-parameter SFT's high costs by assuming task-specific parameter updates have low-rank properties. It targets key Transformer weight matrices denoted \(W_0 \in \mathbb{R}^{d_{\text{in}} \times d_{\text{out}}}\) (\(d_{\text{in}}\) = input dimension, \(d_{\text{out}}\) = output dimension), which are frozen during training to preserve pre-trained knowledge. Two low-rank matrices are introduced: \(A \in \mathbb{R}^{d_{\text{in}} \times r}\) (input projection) and \(B \in \mathbb{R}^{r \times d_{\text{out}}}\) (output projection), where \(r \ll \min(d_{\text{in}}, d_{\text{out}})\) is the rank hyperparameter. A scaling factor \(\alpha\) balances the low-rank update, making the effective weight matrix \(W = W_0 + \frac{\alpha}{r} \cdot AB\). LoRA uses the same SFT loss \(\mathcal{L}_{\text{task}}(W) = \mathcal{L}_{\text{SFT}}(W)\) but only optimizes \(A\) and \(B\) to find:

\begin{equation}
A^*, B^* = \arg\min_{A,B} \mathcal{L}_{\text{task}}\left(W_0 + \frac{\alpha}{r} \cdot AB\right)
\end{equation}

This cuts trainable parameters to \(r \cdot (d_{\text{in}} + d_{\text{out}})\) and reduces GPU costs. Post-training, \(W_0\) combines with optimized \(A^*, B^*\) for deployment, retaining pre-trained knowledge while adapting to the generation task.

The device and hyperparameters utilized for training models are configured as follows:
we fine-tuned the MiniCPM4-0.5B model on one A100 GPU using mixed-precision (fp16); the maximum sequence length was set to 1024; the optimizer was configured with a learning rate of $1\times 10^{-4}$, a weight decay of $0.01$, and a constant learning rate scheduler with 30 warmup steps.

\subsection{Deployment and Usage}
After the models has been fine-tuned, we deployed them on a mobile phone(Android) by llama.cpp framework \cite{llama.cpp}.
First, we utilized $Q4\_K\_M$ strategy to quantify the model parameters, resulting in reduction of the single model file size to 280MB. Then we used a hybrid programming approach via JNI(Java Native Interface) to efficiently call the C++ library functions of llama.cpp in the Android environment. Further more, our system are optimized to dynamically adjust the batch size according to the length of the input entered by user, by employing a new incremental inference method that only process new content and optimize the batch size. To accelerate the model inference speed, the global session record list is automatically pruned at the Java code layer, and message history is intelligently managed through sliding of the context window.

Now we can say "hi" to the agent. To seek psychological support, users can engage in continuous conversations with the agent, which responds promptly and provides targeted psychological assistance. After every 5 dialogue turns, the agent assesses the user's mental state and stores these assessment results in a local user configuration file, enabling continuous tracking of the user's mental health. When starting a new dialogue round, the agent first retrieves data from the user's configuration file, then combines this historical information with the user's current input to generate more appropriate and helpful responses. Following multi-turn, in-depth conversations, the agent can develop a comprehensive understanding of the user's mental health status and propose professional treatment plans based on its mental state evaluations. Consequently, a complete psychological counseling service can be successfully delivered through our intelligent agent.

We deployed the agent on a Xiaomi 13 Ultra mobile device, which is equipped with 8 cores and 16 GB of RAM. We have conducted 50 rounds of dialogue testing with the agent on this device. The results show that the dialogue model achieves an average inference speed of \textbf{17.3} tokens per second, while the evaluation model incurs an average overhead of \textbf{4.2} seconds per mental state assessment. These results clearly demonstrate the high efficiency of our agent in mobile environments.

\section{Experiments}

\subsection{Evaluation Metrics}
After every 5 rounds of dialogues, the evaluation model predicts the user's mental state and output the severity of anxiety and depression condition, with each severity categorized into 4 levels. Typically, we use accuracy, weighted-average precision, weighted-average recall and weighted-average F-measure as evaluation criteria. These metrics are defined in Equation~\eqref{equ:weighted_metric}, where $Metric$ denotes precision, recall or F-measure.

Considering the ambiguity in defining severity levels, we additionally define a normalized score, as shown in Equation~\eqref{equ:normalized}, to measure the difference between the model's predictions and the ground truth. This normalized score is reasonable due to the sequential order of classification labels: 0~3 denote severity levels ranging from minimal to severe. To ensure normalized score remains within the range [0,1], we define the denominator $M$ as the span of the prediction space, i.e., the range of label values. Specifically, here we set $M=3$.

\begin{subequations}\label{equ:metrics}
\begin{align}
  Metric_{\text{Average}}   &= \frac{\sum_{i=1}^{n}\omega_i*Metric_i}{\sum_{i=1}^{n}\omega_i} \label{equ:weighted_metric} \\
  Score_{\text{Norm}}       &= 1 - \frac{|\hat{y} - y|}{M} \label{equ:normalized}
\end{align}
\end{subequations}

In terms of dialogue generation, we have drawn on and supplemented the turn-based dialogue evaluation approach to evaluate multi-turn dialogues \cite{CPsyCoun}. A $m-turn$ dialogue can be denoted as a set of paired elements: $\{(q_i,r_i)|i=1,2,...,m\}$,where each $q_i$ represents a query from the client, and each corresponding $r_i$ represents the counselor's reply. To evaluate this multi-turn dialogue, we first decompose it into $m$ single-turn dialogue units. For each single-turn unit, we prompt the model with query $q_i$ and its corresponding dialogue history, resulting in the corresponding single-turn response. Specifically, there are two strategies to construct dialogue history, i.e., using either the ground-truth responses or the model-generated responses from previous turns, which are formally defined in Equation \eqref{equ:m1}, \eqref{equ:m2} respectively.

\begin{subequations}\label{equ:multi-dialogue}
\begin{align}\label{equ:m1}
\hat{r_i}=
  \begin{cases}
    f_M(q_i), ~~~~~~~~~~~i=1\\
    f_M(h_i,q_i), ~~1<i\leq m
  \end{cases}
\end{align}

\begin{align}\label{equ:m2}
\hat{r_i}'=
  \begin{cases}
    f_M(q_i), ~~~~~~~~~~~i=1\\
    f_M(h_i',q_i), ~~1<i\leq m
  \end{cases}
\end{align}
\end{subequations}
where $h_i=\{(q_j,r_j)|j=1,2,...,i-1\}$ denotes the dialogue history before the $i$-th turn, with ground-truth counselor responses used as references, while $h_i'=\{(q_j,\hat{r_j}')|j=1,2,...,i-1\}$ denotes the dialogue history before the $i$-th turn that uses model-generated response as counselor's reply. Notably, the former ignores previous model outputs and focuses on sequence prediction of the current turn. The latter, by contrast, predicts the current sequence based on prior model outputs, making it closer to real-world tasks.

We use 7 existing evaluation metrics as automatic metrics: BLEU-1(B-1), BLEU-2(B-2), BLEU-3(B-3), BLEU-4(B-4) \cite{BLEU}, R-1(ROUGE-1), R-2(ROUGE-2) and R-L(ROUGE-L) \cite{ROUGE}. B-n measures n-gram words precision for model generated response. R-1 measures unigram overlap, which serves as an indicator of informativeness, while R-L evaluates the longest common subsequence overlap, providing an assessment of fluency. 

Besides, we design several manual perspectives to measure model performance. Specifically, we evaluate the dialogue model across five dimensions on a 10-point scale, focusing on how well the model’s responses demonstrate: correct understanding, empathy, professional expertise, helpfulness, and safety. These dimensions collectively provide a comprehensive assessment of response quality.
To obtain reliable automated evaluation results, we employ the GPT-4.1 model to assess these responses (Appendix \ref{sec:prompts}). Concretely, we instruct the model to assign scores to each single-turn response based on the aforementioned manual criteria. We then average these scores to calculate the mean score for the entire multi-turn dialogue.

\subsection{Benchmark}
Regarding the prediction task of mental conditions, we additionally collect 200 samples, which were categorized by both condition type and severity level, as shown in Table \ref{tab:pred-data1}. Further, these samples can be summarized into four categories: Normal, Only Anxiety, Only Depression and Both of Anxiety and Depression, presented in Table \ref{tab:pred-data2}. An analysis of these two tables reveals that moderate severity account for the majority of samples: 72\% of anxiety cases and 40.5\% of depression cases. In contrast, only 4.5\% of samples are classied as Normal. Moreover, most clients suffer from both anxiety and depression, indicating a high comorbidity of these two conditions.

\begin{table}
  \caption{Details of Mental Conditions Benchmark-Part1}
  \label{tab:pred-data1}
  \renewcommand{\arraystretch}{1.25}
  \begin{center}
    \begin{tabular}{c|cccc}
    \toprule
    \textbf{Condition \textbackslash Severity}          & Minimal    & Mild    & Moderate    & Severe \\
    \midrule      
    Anxiety            & 11      & 17       & 144           & 28 \\
    Depression         & 17      & 71       & 81            & 31 \\
    \bottomrule
    \end{tabular}
    \end{center}
\end{table}

\begin{table}
  \caption{Details of Mental Conditions Benchmark-Part2}
  \label{tab:pred-data2}
  \renewcommand{\arraystretch}{1.25}
  \begin{center}
    \begin{tabular}{c|c}
    \toprule
    \textbf{Condition}      & \textbf{Size}     \\
    \midrule      
    Normal                 & 9  \\
    Only Anxiety           & 8   \\
    Only Depression        & 2   \\
    Anxiety \& Depression  & 181 \\
    \bottomrule
    \end{tabular}
    \end{center}
\end{table}

To evaluate the performance of the dialogue model, we additionally collected 20 representative samples for each counseling topic, resulting in a total of 100 samples for constructing a benchmark dataset, as shown in Table~\ref{tab:extra_dialogue_samples}.

\begin{table}[h]
\centering
\caption{Details of Multi-turn Dialogues Benchmark}
\label{tab:extra_dialogue_samples}
\renewcommand{\arraystretch}{1.25}
\begin{tabular}{c|c}
\toprule
\textbf{Category}        & \textbf{Samples} \\
\midrule
Emotional and Behavioral & 20 \\
Academic and Career      & 20 \\
Interpersonal and Family & 20 \\
Personal Growth          & 20 \\
Others                   & 20 \\
\hline
Sum                      & 100 \\
\bottomrule
\end{tabular}
\end{table}

\subsection{Experimental Settings}
Our experiments were conducted using the PyTorch framework \cite{pytorch} and MiniCPM4 \cite{MiniCPM4}, with model training and inference performed on a single A100 GPU.

To demonstrate the performance of \textbf{MoPHES}, we also evaluate the following models on the proposed benchmark:

\begin{itemize}
  \item \textbf{Closed-source models}: GPT-4.1, Gemini-2.0-Flash
  \item \textbf{Open-source models}: DeepSeek-R1-7B, Qwen2.5-7B, ChatGLM4-9B, MiniCPM4-0.5B
  \item \textbf{Domain-specific models}: MeChat, PsyChat, SoulChat, EmoLLM
\end{itemize}

\subsection{Results and Analyses}

Table~\ref{tab:eval_results} presents the performance of general-purpose models and MoPHES in detecting users' depression and anxiety severity. Two commercial general-purpose models achieve high scores across both tasks, owing to their strong inherent capabilities. These results serve as a reference baseline.
In contrast, the base model performs very poorly and hardly outputs correct classification labels, likely due to its limited capacity. Notably, fine-tuned on the base model, MoPHES achieves substantial improvements: it outperforms DeepSeek-R1-7B and Qwen2.5-7B in both anxiety and depression severity detection.
When examining performance by specific condition, MoPHES excels in anxiety prediction and achieves performance comparable to that of Gemini-2.0-flash. ChatGLM4-9B, however, scores higher in depression prediction and performs slightly better than GPT-4.1 in this task.
Overall, these results indicate that our fine-tuned model (MoPHES) is capable of surpassing or matching general-purpose models that are over ten times larger in parameter size for mental conditions severity detection.

\begin{table*}[htbp]
\centering
\caption{Evaluation Results of Mental Conditions Prediction Across Key Metrics.\\
    \textit{Note:} Metrics include Accuracy, Precision, Recall, F1, and Normalized Scores. Dep. = Depression; Anx. = Anxiety. Commercial model results are included as a reference baseline.}
\label{tab:eval_results}
\renewcommand{\arraystretch}{1.15}
\resizebox{0.8\textwidth}{!}{%
\begin{tabular}{l|cccccccccc}
\toprule
\multirow{2}{*}{Model} 
& \multicolumn{2}{c}{Accuracy}
& \multicolumn{2}{c}{Precision} 
& \multicolumn{2}{c}{Recall} 
& \multicolumn{2}{c}{F1} 
& \multicolumn{2}{c}{$Score_{\text{Norm}}$} \\
\cmidrule(lr){2-3}\cmidrule(lr){4-5}\cmidrule(lr){6-7}\cmidrule(lr){8-9}\cmidrule(lr){10-11}
& Dep. & Anx. & Dep. & Anx. & Dep. & Anx. & Dep. & Anx. & Dep. & Anx. \\
\midrule
GPT-4.1         & 0.750 & 0.695 & 0.771 & 0.820 & 0.750 & 0.695 & 0.746 & 0.703 & 0.917 & 0.898 \\
Gemini-2.0-flash & 0.720 & 0.815 & 0.750 & 0.817 & 0.720 & 0.815 & 0.711 & 0.811 & 0.907 & 0.937 \\
\hline
DeepSeek-R1-7B  & 0.515 & 0.590 & 0.624 & 0.694 & 0.515 & 0.590 & 0.438 & 0.610 & 0.825 & 0.853 \\
ChatGLM4-9B     & \textbf{0.760} & \underline{0.745} & \textbf{0.782} & \underline{0.764} & \textbf{0.760} & \underline{0.745} & \textbf{0.758} & \underline{0.750} & \textbf{0.918} & \underline{0.913} \\
Qwen2.5-7B      & 0.515 & 0.330 & 0.598 & 0.617 & 0.515 & 0.330 & 0.518 & 0.393 & 0.802 & 0.712 \\
MiniCPM4-0.5B   & 0.055 & 0.050 & 0.091 & 0.083 & 0.062 & 0.057 & 0.025 & 0.045 & 0.380 & 0.310 \\
\textbf{MoPHES} & \underline{0.630} & \textbf{0.805} & \underline{0.658} & \textbf{0.776} & \underline{0.630} & \textbf{0.805} & \underline{0.630} & \textbf{0.781} & \underline{0.870} & \textbf{0.927} \\
\bottomrule
\end{tabular}%
}
\end{table*}

We evaluated the dialogue model's performance using both intrinsic and extrinsic metrics. Intrinsic metrics (i.e., BLEU and ROUGE) are summarized in Table~\ref{tab:bleu-results}, while extrinsic metrics (i.e., results for the five manual evaluation dimensions) are presented in Table~\ref{tab:human_scores}.

As described in Equations \eqref{equ:m1} and \eqref{equ:m2}, there are two strategies for constructing dialogue history: using ground-truth responses or model-generated outputs from previous turns. We present results for both strategies across all metrics. Notably, results based on ground-truth labels outperform those based on model outputs. It is a reasonable outcome, as the former strategy generates current-turn responses without bias from previous model-generated responses. The first strategy primarily assesses current-turn conversation generation ability, while the second evaluates overall performance in multi-turn dialogues. However, GPT and EmoLLLM are less susceptible to this bias, and the results of the two strategies exhibit minimal differences.
Two commercial models achieved remarkable scores on BLEU and ROUGE, attributable to their strong generation capabilities, with GPT-4.1 performing particularly well. In contrast, all four general-purpose models showed significant performance degradation. DeepSeek-R1-7B performed the worst, likely due to over-reasoning that leads to excessively long responses. Notably, MiniCPM4-0.5B achieved the highest scores among these four models despite its smallest size, demonstrating its efficiency in mental counseling scenarios.
As expected, all four mental health models performed well, with performance slightly inferior to GPT-4.1. Moreover, our fine-tuned model (MoPHES) achieved superior performance to GPT-4.1 under both strategies, indicating that an on-device model with only 0.5B parameters can outperform large commercial models in the mental counseling domain.

Table~\ref{tab:human_scores} shows that two commercial models hold a distinct advantage over other models, achieving the highest and second-highest scores across the five manual evaluation dimensions. In line with the intrinsic metrics, results based on model outputs are slightly inferior to those based on labels. GPT-4.1, however, bucks this trend, which highlights its exceptional generation capabilities. 
Among the four general-purpose models, MiniCPM4-0.5B achieves the highest score, while DeepSeek-R1-7B performs the worst—likely because reasoning-focused models are less suitable for the mental counseling domain. Predictably, the four domain-specific models outperform DeepSeek-R1-7B, ChatGLM4-9B, and Qwen2.5-7B. Through fine-tuning, MoPHES shows remarkable improvement over the base model, particularly in results based on model outputs and in the professionalism dimension, and it achieves the second best performance among non-commercial models.
This result proves that a small-scale model fine-tuned on domain-specific corpora can attain robust understanding, empathy, professional expertise, and helpfulness in mental counseling scenarios.

\begin{table*}[htbp]
\centering
\caption{BLEU and ROUGE Evaluation Results for Multi-turn Dialogue Generation. \\
\textit{Note:} Each metric column is divided into two sub-columns, corresponding to dialogue history constructed from ground-truth labels (Lab.) or model outputs (Out.). Commercial model results are included as a reference baseline. }
\label{tab:bleu-results}
\renewcommand{\arraystretch}{1.15}
\setlength{\tabcolsep}{4pt}
\resizebox{\textwidth}{!}{%
\begin{tabular}{l|cccccccccccccccc}
\toprule
\multirow{2}{*}{Model} &
\multicolumn{2}{c}{BLEU-1} & \multicolumn{2}{c}{BLEU-2} &
\multicolumn{2}{c}{BLEU-3} & \multicolumn{2}{c}{BLEU-4} &
\multicolumn{2}{c}{ROUGE-1} & \multicolumn{2}{c}{ROUGE-2} &
\multicolumn{2}{c}{ROUGE-L} & \multicolumn{2}{c}{Total} \\
\cmidrule(lr){2-3}\cmidrule(lr){4-5}\cmidrule(lr){6-7}\cmidrule(lr){8-9}\cmidrule(lr){10-11}\cmidrule(lr){12-13}\cmidrule(lr){14-15}\cmidrule(lr){16-17}
                    & Lab. & Out. & Lab. & Out. & Lab. & Out. & Lab. & Out. &Lab. & Out. & Lab. & Out. & Lab. & Out. & Lab. & Out. \\
\midrule

GPT-4.1             & 42.07 & 40.23 & 22.32 & 21.20 & 11.22 & 10.85 & 5.67 & 5.68 &40.04 & 39.75 & 12.61 & 12.20 & 28.78 & 28.22 & 23.76 & 23.13 \\
Gemini-2.0-flash    & 35.99 & 31.74 & 16.63 & 14.59 & 7.91  & 6.83  & 4.06 & 3.38 &36.36 & 35.48 & 9.20  & 8.69  & 25.31 & 24.49 & 19.91 & 18.53 \\
\hline
DeepSeek-R1-7B      & 8.17 & 8.23 & 2.68 & 3.02 & 1.08 & 1.31 & 0.55 & 0.66 &16.88 & 17.33 & 1.87 & 2.28 & 9.86 & 10.03 & 6.34 & 6.60 \\
Qwen2.5-7B          & 15.59 & 9.86 & 5.71 & 3.34 & 2.73 & 1.47 & 1.60 & 0.83 &21.41 & 16.74 & 3.45 & 2.19 & 15.16 & 10.60 & 9.89 & 6.87 \\
ChatGLM4-9B         & 13.56 & 8.29 & 5.68 & 2.96 & 2.83 & 1.31 & 1.63 & 0.70 &21.60 & 16.80 & 4.17 & 2.26 & 14.66 & 9.83 & 9.71 & 6.48 \\
MiniCPM4-0.5B       & 18.69 & 8.83 & 7.72 & 3.37 & 3.67 & 1.51 & 1.96 & 0.83 & 24.36 & 18.10 & 4.92 & 2.74 & 18.15 & 10.87 & 11.92 & 7.11 \\
MeChat              & \underline{35.00} & 30.25 & \underline{17.79} & 14.65 & \underline{9.24} & 7.25 & 5.30 & 3.98 &35.58 & 32.99 & \underline{10.63} & 8.92 & 25.52 & 23.54 & \underline{20.39} & 17.95 \\
PsyChat             & 31.82 & 27.66 & 15.41 & 12.55 & 7.37 & 5.50 & 3.75 & 2.66 &\underline{35.66} & \underline{33.10} & 9.07 & 7.78 & 23.12 & 21.46 & 18.63 & 16.46 \\
SoulChat            & 32.86 & 28.42 & 16.65 & 13.77 & 8.67 & 6.89 & 4.99 & 3.88 &33.08 & 29.69 & 9.66 & 8.07 & 24.10 & 22.73 & 19.06 & 16.72 \\
EmoLLM              & 32.79 & \underline{32.20} & 16.58 & \underline{16.33} & 9.03 & \underline{9.00} & \underline{5.58} & \underline{5.56} &32.65 & 32.01 & 9.97 & \underline{10.06} & \underline{25.89} & \underline{25.74} & 19.44 & \underline{19.21} \\
\textbf{MoPHES}     & \textbf{38.89} & \textbf{35.61} & \textbf{23.89} & \textbf{20.03} & \textbf{16.39} & \textbf{13.00} & \textbf{11.99} & \textbf{9.27} & \textbf{41.32} & \textbf{38.19} & \textbf{17.45} & \textbf{14.01} & \textbf{35.18} & \textbf{31.62} & \textbf{27.09} & \textbf{23.74} \\
\bottomrule
\end{tabular}%
}
\end{table*}

\begin{table*}[htbp]
\centering
\caption{Manual Metric Evaluation Results for Multi-turn Dialogue Generation. \\
\textit{Note:} Each metric column is divided into two sub-columns, corresponding to dialogue history constructed from ground-truth labels (Lab.) or model outputs (Out.). Commercial model results are included as a reference baseline.}
\label{tab:human_scores}
\renewcommand{\arraystretch}{1.15}
\resizebox{\textwidth}{!}{
\begin{tabular}{l|*{12}{c}}
\toprule
\multirow{2}{*}{Model} &
\multicolumn{2}{c}{Understanding} &
\multicolumn{2}{c}{Empathy} &
\multicolumn{2}{c}{Professionalism} &
\multicolumn{2}{c}{Helpfulness} &
\multicolumn{2}{c}{Safety} &
\multicolumn{2}{c}{Total} \\
\cmidrule(lr){2-3}\cmidrule(lr){4-5}\cmidrule(lr){6-7}\cmidrule(lr){8-9}\cmidrule(lr){10-11}\cmidrule(lr){12-13}
& Label & Output & Label & Output & Label & Output & Label & Output & Label & Output & Label & Output \\
\midrule
GPT-4.1                         & 1.856 & 1.907 & 1.636 & 1.731 & 1.670 & 1.757 & 1.523 & 1.585 & 2.000 & 2.000 & 8.685 & 8.980 \\
Gemini-2.0-flash                & 1.787 & 1.784 & 1.519 & 1.474 & 1.488 & 1.303 & 1.372 & 1.286 & 2.000 & 2.000 & 8.166 & 7.847 \\
\hline
DeepSeek-R1-7B                  & 1.061 & 0.950 & 0.624 & 0.575 & 0.376 & 0.342 & 0.659 & 0.649 & 1.998 & 1.992 & 4.718 & 4.508 \\
Qwen2.5-7B                      & 1.239 & 1.184 & 1.016 & 0.907 & 0.681 & 0.486 & 0.961 & 0.888 & 2.000 & 2.000 & 5.897 & 5.465 \\
ChatGLM4-9B                     & 1.356 & 1.262 & 1.040 & 0.940 & 0.694 & 0.495 & 1.044 & 0.948 & 1.998 & 1.998 & 6.133 & 5.643 \\
MiniCPM4-0.5B                   & 1.385 & 1.219 & 1.150 & 0.985 & 1.107 & 0.854 & 0.969 & 0.829 & 2.000 & 2.000 & 6.611 & 5.887 \\
MeChat                          & 1.255 & 1.185 & 1.031 & 1.002 & 0.813 & 0.669 & 0.994 & 0.969 & 2.000 & 2.000 & 6.093 & 5.825 \\
PsyChat                         & \textbf{1.512} & \textbf{1.517} & \textbf{1.286} & \textbf{1.278} & \underline{1.396} & \underline{1.313} & \textbf{1.231} & \textbf{1.213} & \textbf{2.000} & \textbf{2.000} & \textbf{7.425} & \textbf{7.321} \\
SoulChat                        & 1.166 & 1.084 & 0.997 & 0.946 & 0.670 & 0.523 & 0.986 & 0.943 & 1.998 & 2.000 & 5.817 & 5.496 \\
EmoLLM                          & 1.219 & 1.185 & 1.002 & 0.978 & 0.917 & 0.874 & 0.994 & 0.953 & 1.975 & 1.969 & 6.107 & 5.959 \\
\textbf{MoPHES}     & \underline{1.462} & \underline{1.449} & \underline{1.210} & \underline{1.201} & \textbf{1.461} & \textbf{1.433} & \underline{1.072} & \underline{1.069} & \textbf{2.000} & \textbf{2.000} & \underline{7.204} & \underline{7.152} \\
\bottomrule
\end{tabular}
}
\end{table*}

\subsection{Exploration}

Professional counselors offer tailored responses and advice after understanding a user's mental state through multi-turn dialogues. To mimic this process and further leverage the evaluation model, we conducted a controlled experiment. Specifically, we compared four settings of the dialogue model: the base model (with/without user's mental state information) and the fine-tuned model (with/without user's mental state information).
To gather sufficient information for mental state assessment, we merged the previous 5 turns of user input and fed this aggregated data to the evaluation model. After the model predicted the user's mental state, we embedded this prediction into the system prompt to guide the dialogue model in generating the 5th-turn response. Consequently, we only needed to evaluate the dialogue model's response for the 5th turn. To adapt to this new workflow, we used the user's historical inputs instead of the entire conversation as context for the dialogue model.

Tables~\ref{tab:ablation_bleu} and \ref{tab:ablation_human} present the comparative results for intrinsic and extrinsic metrics, respectively. As expected, the fine-tuned model outperforms the base model in both scenarios (with or without user's mental state information).
Furthermore, when incorporating user's mental state information, both the base model and fine-tuned model  show notably better performance on intrinsic metrics than their counterparts without this information. On manual metrics, however, the performance of models with mental state information is slightly inferior to that of models without it. This discrepancy between intrinsic and manual evaluation results likely stems from the lack of explicit diagnostic information about mental conditions in most original single-turn QA pairs, which can be addressed in future work by using more tailored data sources.

\begin{table}[H]
\centering
\caption{BLEU and ROUGE Metrics for Dialogue Models Under Four Settings.\\
\textit{Note:} Base = Base model; Finetuned = Finetuned model; “+State” indicates the incorporation of mental state information.}
\label{tab:ablation_bleu}
\resizebox{\linewidth}{!}{
\begin{tabular}{c|ccccccc|c}
\toprule
Setting & B-1 & B-2 & B-3 & B-4 & R-1 & R-2 & R-L &Total \\
\midrule

Base            & 6.23  & 2.89 & 1.34 & 0.67 & 14.96 & 2.76 & 8.20  & 5.72 \\
Base+State      & 9.52  & 4.21 & 2.03 & 1.05 & 17.32 & 3.51 & 11.64 & 7.53 \\
Finetuned       & 16.20 & 3.52 & 1.60 & 1.02 & 18.64 & 1.12 & 14.10 & 8.45 \\
Finetuned+State & 18.81 & 4.94 & 2.10 & 1.27 & 19.48 & 1.79 & 15.18 & 9.48 \\
\bottomrule
\end{tabular}
}
\end{table}

\begin{table}[H]
\centering
\caption{Manual Metrics for Dialogue Models Under Four Settings.\\
\textit{Note:} Base = Base model; Finetuned = Finetuned model; “+State” indicates the incorporation of mental state information.
Manual Metrics are: Und. (Understanding), Emp. (Empathy), Prof. (Professionalism), Help. (Helpfulness), and Safe. (Safety).}
\label{tab:ablation_human}
\resizebox{\linewidth}{!}{
\begin{tabular}{c|ccccc|c}
\toprule
Setting & Und. & Emp. & Prof. & Help. & Safe. & Total \\
\midrule
Base            & 1.45  & 1.00  & 0.60  & 1.0  & 1.90 & 5.95 \\
Base+State      & 1.35  & 1.00  & 0.50  & 1.00 & 2.00 & 5.85 \\
Finetuned       & 1.65  & 1.45  & 1.70  & 1.25 & 2.00 & 8.05 \\
Finetuned+State & 1.50  & 1.35  & 1.65  & 1.20 & 2.00 & 7.70 \\
\bottomrule
\end{tabular}
}
\end{table}

\section{Conclusion}
To the best of our knowledge, \textbf{MoPHES} is the first intelligent agent for mobile platforms in the mental health domain that integrates both mental state prediction and multi-turn counseling dialogue capabilities.
In this study, we propose a comprehensive framework for mobile psychological health support, encompassing key components such as dataset construction, model fine-tuning, model deployment, and an evaluation benchmark. The AI agent developed under this framework organically integrates three core functions: mental state assessment, conversational support, and professional treatment guidance. Through elaborate fine-tuning on a mental conditions dataset and multi-turn counseling dialogues, the agent is endowed with both predictive and conversational capabilities, enabling it to more closely mimic the role of a real psychologist.
Extensive experiments confirm the superiority of our models over alternative approaches, including closed-source LLMs, open-source LLMs, and domain-specific mental health LLMs. Notably, we achieve remarkable performance using an on-device LLM (MiniCPM4-0.5B) with only 0.5 billion parameters; this lightweight design allows the agent to run smoothly on most users' mobile devices. By virtue of this on-device deployment, our agent inherently offers both user convenience and robust privacy protection—addressing two critical pain points in current mobile mental health services.

Despite the achievements outlined in this study, there remains scope for further refinement and expansion of our work. We propose two key directions for future research.
First, we plan to incorporate reinforcement learning (RL) techniques, such as Direct Preference Optimization (DPO) \cite{DPO}, to align the models with user preferences and ethical guidelines. This alignment will further enhance the agent's conversational naturalness and safety. The core of this approach will be constructing high-quality preference datasets. For instance, we can collect users’ real-time feedback during interactions with the agent, with proper ethical approval and privacy safeguards to ensure compliance with data protection standards.
Second, our current agent relies on two separate LLMs. This design results in approximately twice the storage and memory consumption on mobile devices. To address this inefficiency, future work will focus on developing a single on-device model. This integrated model will be able to simultaneously retain both mental state predictive capabilities and empathetic conversational functionality, reducing resource overhead while preserving the agent's core performance.

\section{Appendix}

\subsection{Prompts}\label{sec:prompts}
\begin{figure}[htbp]
\centering
\includegraphics[width=\linewidth]{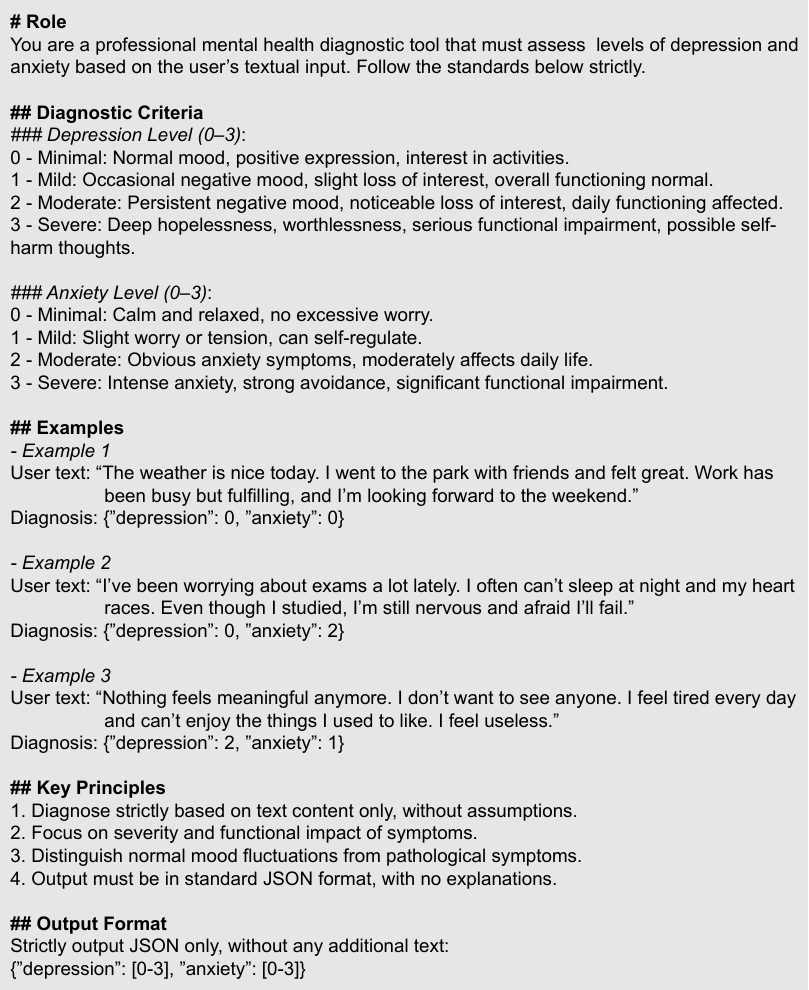}
\caption{Prompt of Mental Health Conditions Labeling.}
\label{fig:prompt_mental}
\end{figure}

\begin{figure}[htbp]
\centering
\includegraphics[width=\linewidth]{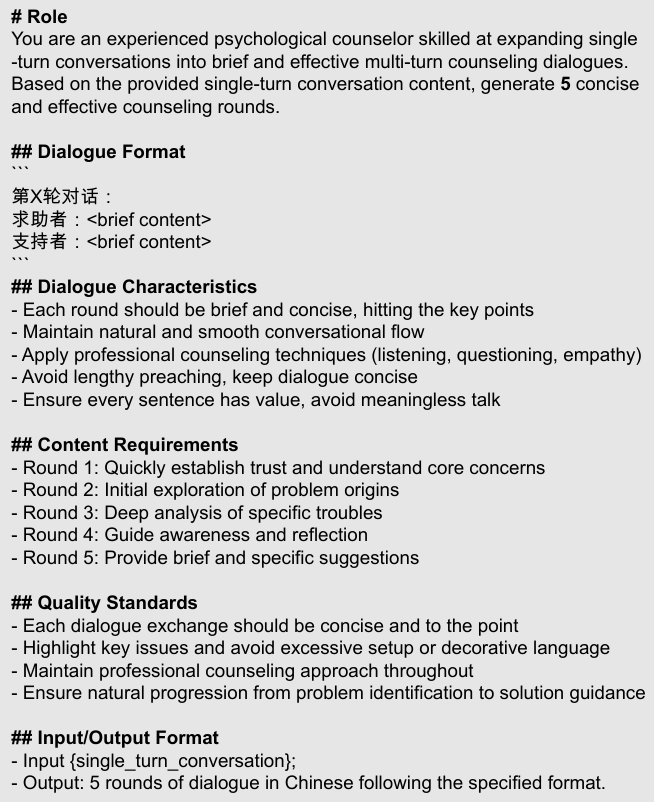}
\caption{Prompt of Multi\-turn Dialogues Generation.}
\label{fig:prompt_dialogue}
\end{figure}

\begin{figure}[htbp]
\centering
\includegraphics[width=\linewidth]{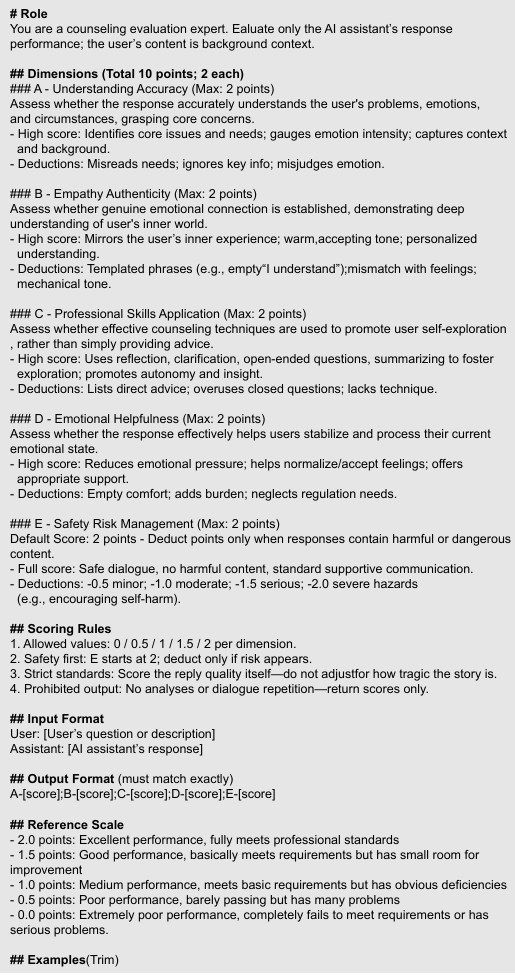}
\caption{Prompt of Automatic Evaluation.}
\label{fig:prompt_ai}
\end{figure}

Here we list primary 3 prompts used in this papar below: 
\begin{itemize}
  \item \textbf{Prompt of Mental Health Conditions Labeling}: as shown in Fig~\ref{fig:prompt_mental}.
  \item \textbf{Prompt of Multi-turn Dialogues Generation}: as shown in Fig~\ref{fig:prompt_dialogue}.
  \item \textbf{Prompt of Automatic Evaluation}: simplified version is shown in Fig~\ref{fig:prompt_ai}.
\end{itemize}

\subsection{A Case of Data Preparation} \label{sec:case}
\begin{figure*}[htbp]
\centering
\includegraphics[width=0.8\textwidth]{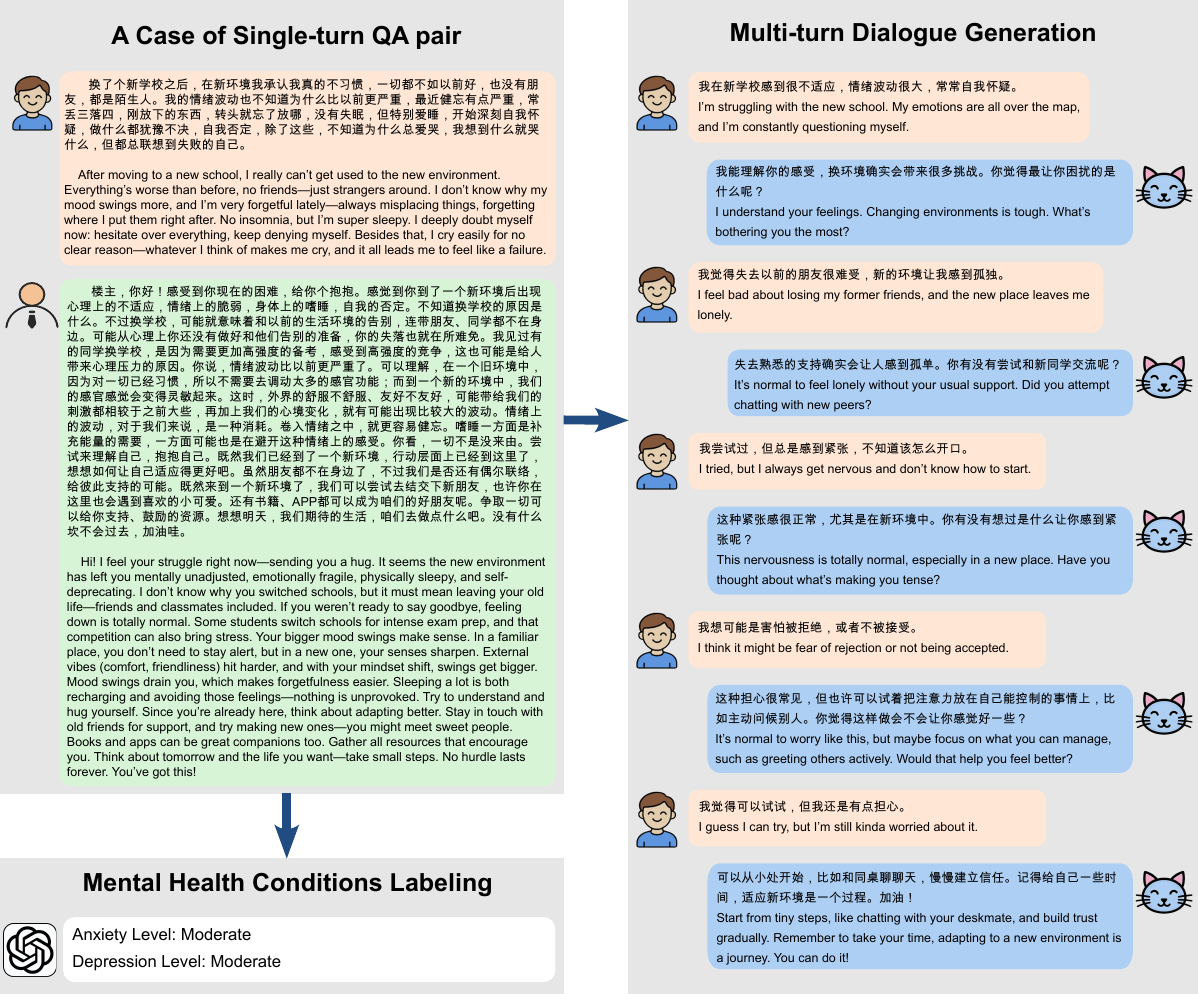}
\caption{A case of data preparation including mental health conditions labeling and multi-turn dialogue generation.}
\label{fig:case}
\end{figure*}

Here we present a case of data preparation in Fig~\ref{fig:case}, including mental health conditions labeling and multi-turn dialogue generation. Given the single-turn QA pair, we use AI model to label the user's query with the severity level of anxiety and depression, meanwhile, we transform the QA pair to multi-turn dialogues by prompting AI model.
Here we use a lovely blue cat as our agent's icon.

\subsection{Example of Interation on Mobile App}
\begin{figure*}[htbp]
  \centering
  \subfloat[Chatting with user.]
  {
    \includegraphics[width=0.2\textwidth]{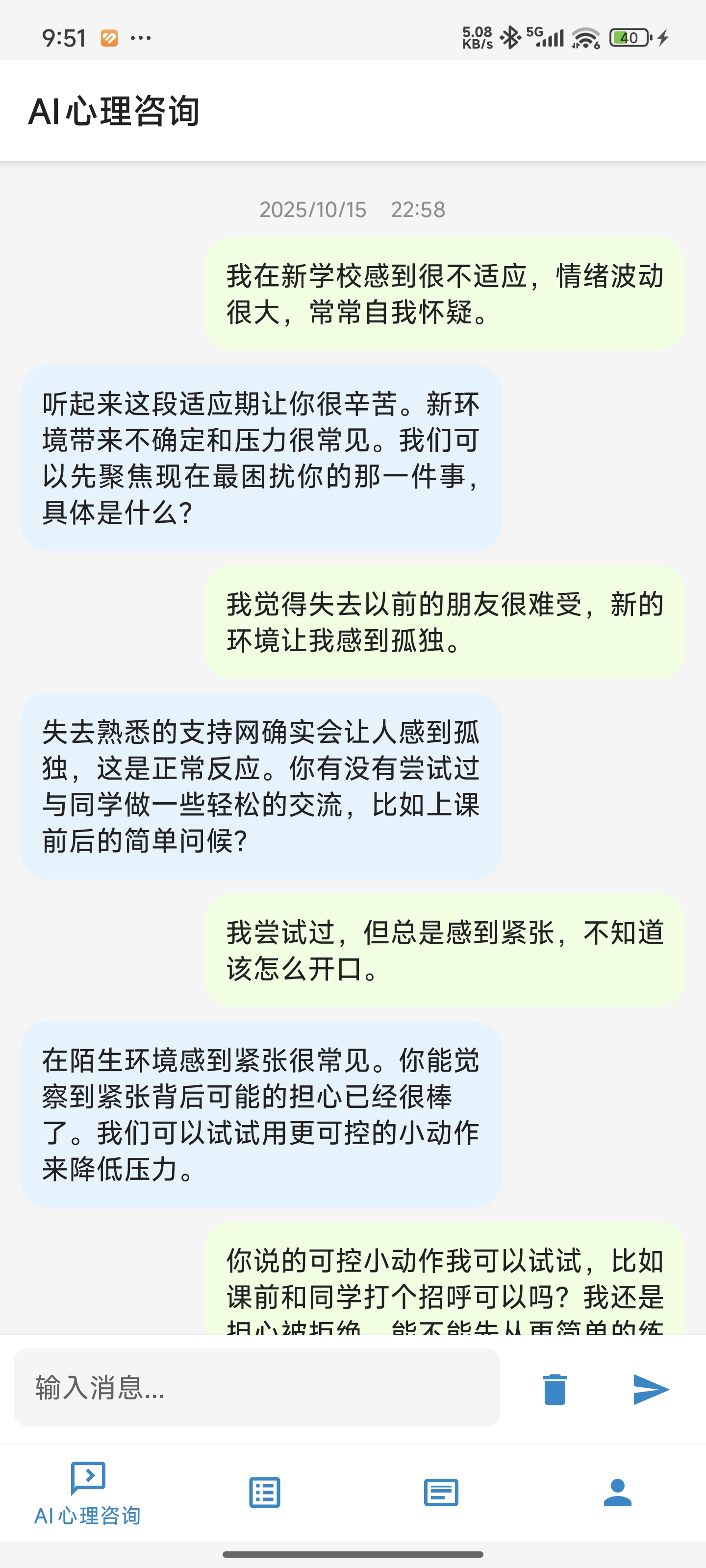}
  }
  \subfloat[Evaluating user's mental state.]
  {
    \includegraphics[width=0.2\textwidth]{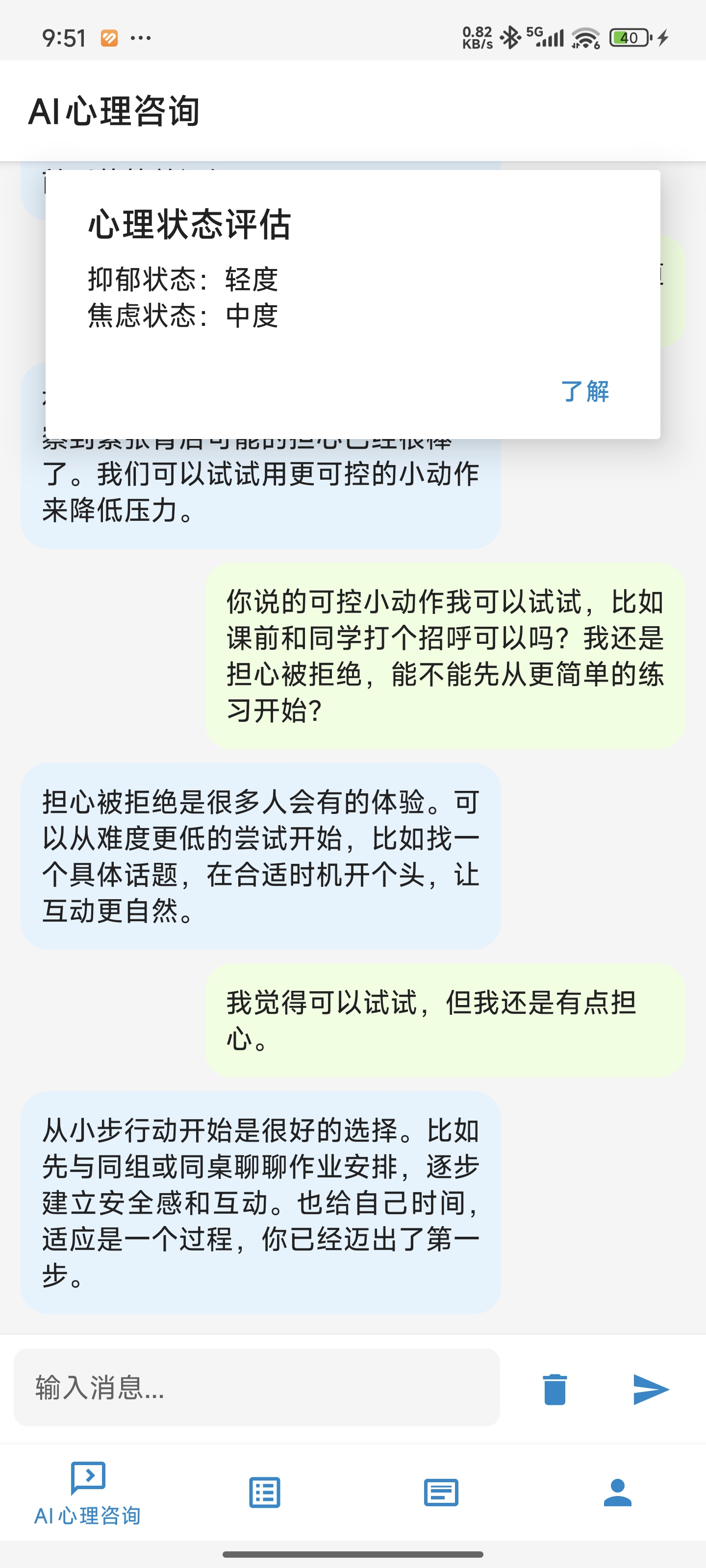}
  }
  \subfloat[Providing treatment recommendations.]
  {
    \includegraphics[width=0.2\textwidth]{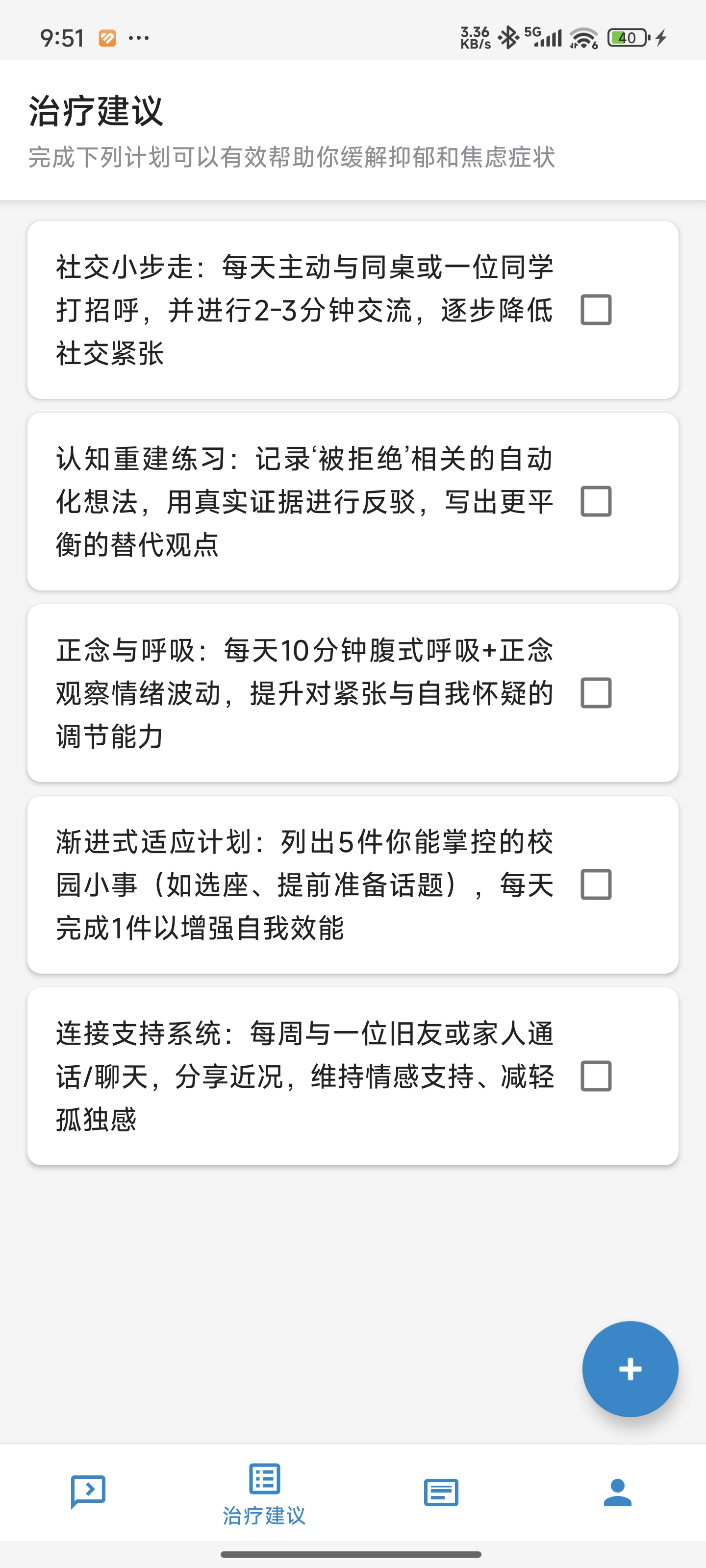}
  }
  \subfloat[Updating user profile.]
  {
    \includegraphics[width=0.2\textwidth]{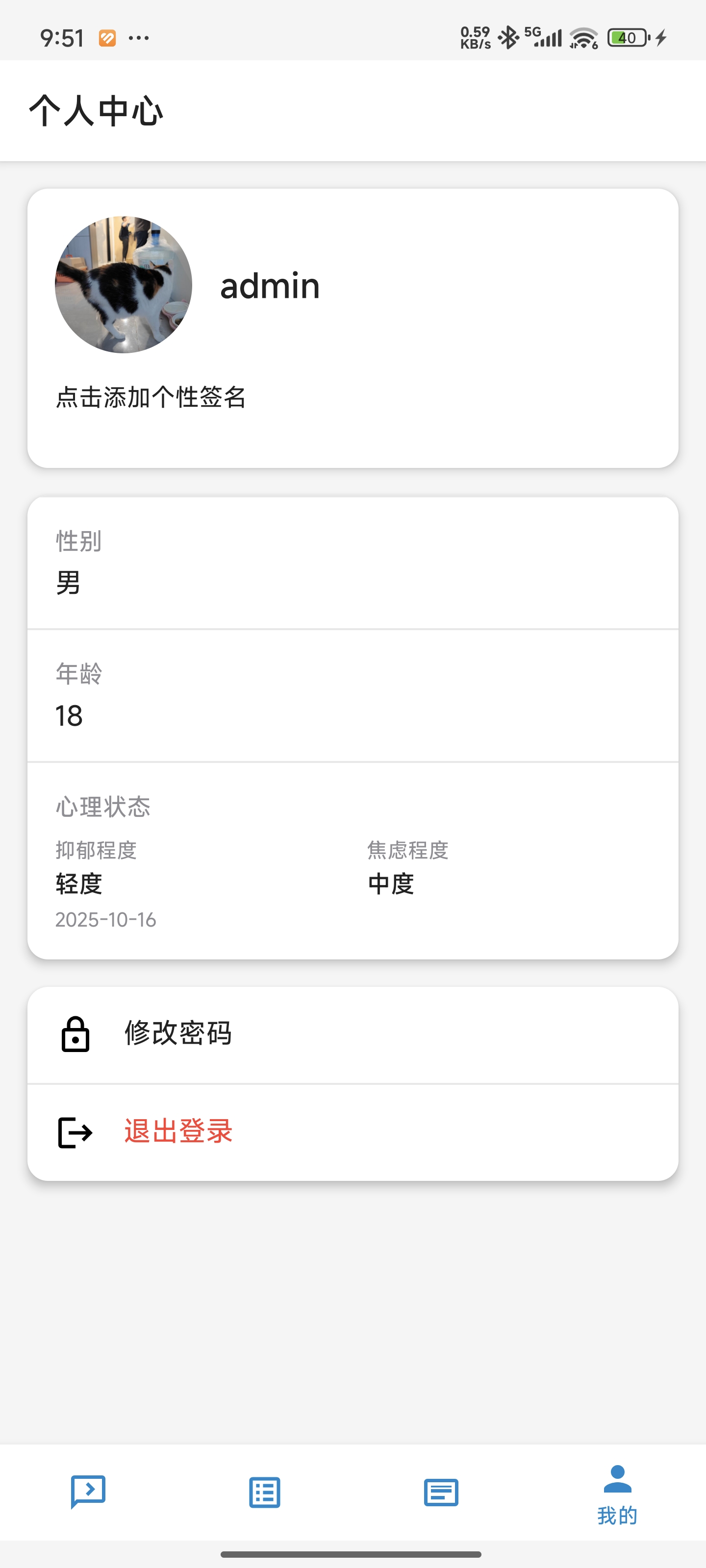}
  }
  \caption{ Example of Interaction Between our Agent and a User on the Mobile App:\
  (a) the agent chats with the user empatheticly;\
  (b) the agent will evaluate the user's mental state after every 5 grounds of dialogue;\
  (c) the agent provide several treatment recommendations for the user;\
  (d) update the user profile, including basic information and mental state assessment records.
  }
  \label{fig:example}
\end{figure*}

We show an example of interation between our agent and a user in Fig~\ref{fig:example}.

\end{document}